\documentclass[12pt,a4paper]{article}
\usepackage{amsmath}
\usepackage[dvips]{graphicx}
\input epsf
\usepackage{times}
\begin{document}
\begin{titlepage}
\title{On the flat transverse momentum dependence of the single-spin asymmetry  in  inclusive neutral pion production}
\author{S.M. Troshin,
 N.E. Tyurin\\[1ex]
\small  \it Institute for High Energy Physics,\\
\small  \it Protvino, Moscow Region, 142281, Russia} \normalsize
\date{}
\maketitle

\begin{abstract}
We discuss recent experimental results from RHIC where the flat 
transverse momentum dependence of a single-spin asymmetry has been found  
in  the inclusive  production of neutral  pions. This dependence takes place in a wide region
of the transverse momenta up to $p_T=10$ GeV/c. We emphasize that similar dependence has been predicted in
the nonperturbative spin filtering mechanism for the
single-spin asymmetries in hadron interactions and present some implications for this mechanism from
the new experimental data.
\end{abstract}

\end{titlepage}
\setcounter{page}{2}

\section*{Introduction}

It should be noted that
decrease  with transverse momentum $p_T$ of  single--spin asymmetries (SSA)  is a common feature of perturbative QCD approaches including 
those based on  account for the various modifications implemented into the calculation scheme originally 
grounded on collinear factorization. The most recent progress in this field is described in \cite{yura,metz}.
The decreasing dependence  has not  been directly observed 
experimentally, but the experimental data were not very conclusive due to large statistcal errors.  However, the existing 
data are consistent with the flat transverse dependence of SSA in inclusive processes. This conclusion is valid for the old data on
 of $\Lambda$--hyperon polarization \cite{lambda}, for example, and for the most recent data obtained at RHIC \cite{fpio}. 
It is essential that the new data cover  the wide region of the
transverse momentum values up to $p_T\simeq 10$ GeV/c.  
In \cite{ttech} it was noted  that  experimental data at higher values of $p_T$ 
 would be needed  to perform a more conclusive test of various pQCD theoretical approaches
and their predictions. The model proposed in \cite{ttech} provided flat $p_T$ -dependence of SSA. 
The new  data which have appeared very recently\cite{fpio, fpio1} 
are consistent with the observed earlier trend, i.e.
flat dependence on transverse momenta can be extended to the region of  higher values of $p_T$.
 Since such high values of transverse momenta have been 
reached experimentally, one forced to conclude that the mechanism of the
 SSA's  generation can have a nonperturbative origin.  Of course, presence of the significant statistical errors in the current 
experimental data (cf. e.g   \cite{fpio,fpio1})
is the serious obstacle on the way of derivation of  a completely unambiguous final conclusion on the impossibility of a decreasing dependence of 
the SSA with $p_T$.

\section*{Highlights  of the filtering mechanism of SSA generation and new large-$p_T$ experimental data}
A nonperturbative  QCD  dynamics is  closely interrelated with the two well--known
phenomena, namely, color confinement  and spontaneous breaking of chiral symmetry   ($\chi$SB)( cf. e.g. \cite{mnh}).
The $\chi$SB--mechanism  resulting in 
transition of current into  constituent quarks  is directly responsible
for generation of their masses and appearance of quark condensates.
  Constituent quarks are colored objects, they appear to be  quasiparticles and a 
hadron  is often represented as a loosely bounded system of the
constituent quarks.
Simultaneously,
  the Goldstone bosons which are the excitations of the condensates appear and mediate interactions of the constituent quarks. 

This interaction is mainly due to a pseudoscalar  pion field  and has therefore
a spin--flip nature.
  Spin states filtering results from the  unitarization process in the $s$-channel and 
connects SSA with  asymmetries in the
position (impact parameter) space \cite{ttech}.

The common features of SSA measurements at RHIC and Tevatron (linear
increase of asymmetry with $x_F$ and flat transverse momentum dependence at large transverse momentum, $p_T>1$ GeV/c)
 are  reproduced and  described
    in the framework of the  semiclassical picture based
   on  the further development
   of the chiral quark model suggested in \cite{csn} and results of its use 
   for the treatment of the polarized and unpolarized inclusive cross-sections including those obtained at RHIC \cite{star}.
      
We summarize now  the essential features of the mechanism. Valence constituent quarks   are
 scattered simultaneously (due to strong coupling with Goldstone bosons)
and in a quasi-independent way by the effective strong
 field. In the initial state of the reaction $pp_\uparrow\to \pi^0 X$ the proton is polarized
 and for the wave function of the proton we use the simple SU(6) model. 
The constituent quark $Q_\uparrow$
with transverse up spin   fluctuates into Goldstone boson and another constituent quark $Q'_\downarrow$ with down spin,
  performing  spin-flip transition \cite{cheng}:
\begin{equation}\label{trans}
Q_\uparrow\to GB+Q'_\downarrow.
\end{equation}
It should be noted that $\pi^0$-fluctuation of constituent quark does not change its flavor and color.
Assuming the equal probabilities of the processes with $U$ and $D$ quarks:
\begin{equation}\label{transu}
U_{\uparrow,\downarrow}\to \pi^0+U_{\downarrow,\uparrow}
\quad
\mbox{and}
\quad
D_{\uparrow,\downarrow}\to \pi^0+D_{\downarrow,\uparrow},
\end{equation}
the production of $\pi^0$ by the polarized proton $p_\uparrow$ in this simple $SU(6)$
picture can be treated as
a result of the fluctuation of the constituent quark $Q_\uparrow$ ($Q=U$ or $D$) in the effective
field into the system $\pi^0+Q_{\downarrow}$ (Fig. 1).
\begin{figure}[h]
\begin{center}
  \resizebox{6cm}{!}{\includegraphics*{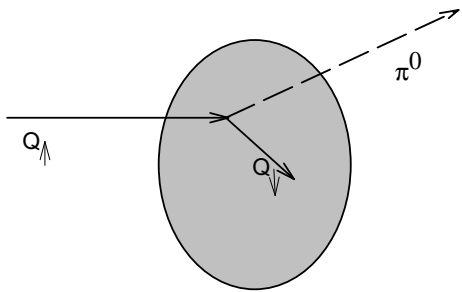}}
\end{center}
\caption{Chiral mechanism of $\pi^0$--production in polarized proton-proton interaction.
 \label{ts1}}
\end{figure}
Since the total angular momentum is conserved, the compensation of quark spin flip should occur, i.e.
to compensate quark spin flip $\delta {\bf S}$, an orbital angular momentum
$\delta {\bf L}=-\delta {\bf S}$ should be associated with the final state of reaction (\ref{trans}).
The introduction of $\delta {\bf L}$  implies
a shift in the impact parameter
value of the pion $\pi^0$:
\[
\delta {\bf S}\Rightarrow\delta {\bf L}\Rightarrow\delta\tilde{\bf b}.
\]
Note, that outside the hadron interior the Goldstone bosons are the usual pions and kaons.
Due to   different strengths of interaction at the different
impact distances, i.e.
\begin{eqnarray}
\nonumber p_\uparrow\Rightarrow   Q_\uparrow & \to & \pi^0 + Q_\downarrow\Rightarrow\;
\;-\delta\tilde{\bf b}, \\
\label{spinflip}p_\downarrow\Rightarrow Q_\downarrow & \to & \pi^0 + Q_\uparrow\Rightarrow\;
\;+\delta\tilde {\bf b}.
\end{eqnarray}
the processes of transition $Q_\uparrow$ and $Q_\downarrow$ to $\pi^0$
 will have different probabilities. It eventually leads  to nonzero asymmetry
$A_N(\pi^0)$.
When the shift in impact
parameter is negative, $-\delta\tilde {\bf b}$, the
interaction is stronger than that with the positive shift,  $+\delta\tilde {\bf b}$,
and therefore the asymmetry $A_N(\pi^0)$ is positive too.
The shift in $\tilde{\bf b}$
(the impact parameter of final pion)
is correlated with the shift of the impact parameter of the collision according
to the relation between impact parameters in the multiparticle production process:
\begin{equation} \label{bi}
{\bf b}=\sum_i x_i{ \tilde{\bf  b}_i}.
\end{equation}
The production of $\pi^0$ with impact parameter $\tilde {\bf b}$ is considered in the fragmentation region, i.e.
at large $x_F$  the approximate relation
\begin{equation}\label{bx}
{\bf b}\simeq x_F\tilde {\bf b},
\end{equation}
which results from Eq. (\ref{bi}) has been used with  additional assumption on the
small values of Feynman $x_F$ for the other particles. 

It should be noted that direction of spin of the proton is chosen to be along or opposite to the impact parameter vector ${\bf b}$. The integration over
the asimuthal angle $\varphi$ selects therefore the finctions $I$ (see below) corresponding to the respective directions of the proton spin.
The mechanism is illustrated in Fig. 2.
\begin{figure}[h]
\begin{center}
  \resizebox{12cm}{!}{\includegraphics*{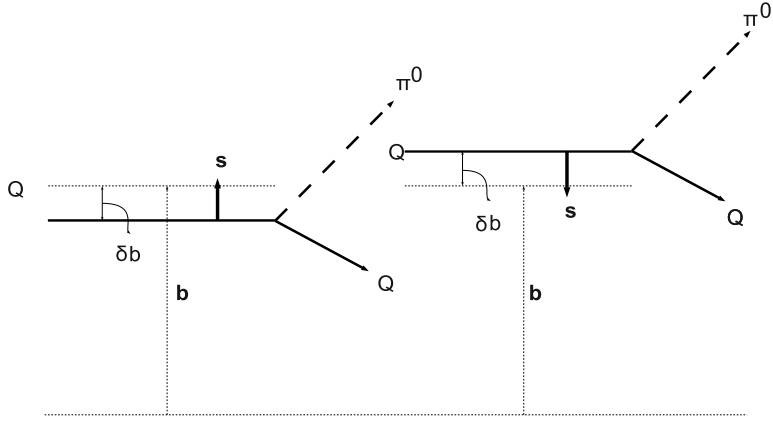}}
\end{center}
\caption{Schematic illustration of the SSA's generation mechanism due to   spin   flip of the constituent quark Q.
 \label{ts2}}
\end{figure}

An essential feature of the mechanism is an account of
unitarity in the direct channel of
reaction. 
 The corresponding formulas for inclusive
cross--sections of the process
\[ p^{\uparrow,\downarrow}+p\to \pi^0 +X, \] have been obtained in
\cite{tmf}:
\begin{equation}
{d\sigma^{\uparrow,\downarrow}}/{d\xi}= 8\pi\int_0^\infty
bdb{I^{\uparrow,\downarrow}(s,b,\xi)}/ {|1-iU(s,b)|^2},\label{un}
\end{equation}
$b$ is the collison impact  parameter.  The function
$U(s,b)$ is the generalized reaction matrix (averaged over initial spin states)
which is determined by the basic dynamics of the elastic scattering.
 The elastic scattering amplitude in the impact
parameter representation $F(s,b)$
   is then given \cite{csn}
 by the  relation:
  \begin{equation} F(s,b)=U(s,b)/[1-iU(s,b)].
\label{6} \end{equation}
The equation (\ref{6}) allows one to obey unitarity for the elastic scattering amplitude provided the inequality
  $ \mbox{Im}\,U(s,b)\geq 0\,$  takes place. The model \cite{csn} has been used for construction
of the functional dependence of  $U(s,b)$, namely, this function was chosen as a product of the factors 
corresponding to the averaged amplitudes of the individual valence quarks. The strong interaction radius
of the quarks is determined by its mass,
\[
 r_Q=\zeta/m_Q
\]
The parameter $\zeta$ was extracted from the experimental data for the differential cross-section
of the elastic $pp$-scattering. In the region of medium values of $t$ this model provides\cite{csn} the familiar Orear-type
behavior:
\[
 \frac{d\sigma}{dt}\sim \exp \left(-\frac{2\pi\zeta}{M}\sqrt{-t}\right),
\]
where $M$ is equal
 to the total mass of the constituent quarks in the two colliding protons, i.e. $M=6m_Q\simeq 2$ GeV/c, and
 the value of parameter $\zeta \simeq 2$ since from the experimental data $m_Q/\zeta=150-200$ MeV
and to reproduce the standard constituent quark masses the value of $\zeta$ should be around 2.

The functions $I^{\uparrow,\downarrow}$ in Eq. (\ref{un}) can be expressed through  the
functions   $U_n^{\uparrow,\downarrow}$ --
  the multiparticle
analogs of the function $U$ \cite{tmf} in the polarized case.
The set of the kinematical variables $\xi$
($x_F$ and $p_T$ for example) describe the state of the produced pion.
   
We assume that the shift $\delta\tilde b$ can be connected with the radius of quark interaction
$r_{Q}^{flip}$, which is
responsible for the quark transition flipping its  spin, i.e.:
\[
\delta\tilde b\simeq r_{Q}^{flip}.
\]
Asymmetry  $A_N$
can be written in terms of the functions $I_{-}$, $I_{0}$ and $U$:
\begin{equation} A_N(s,\xi)=\frac{\int_0^\infty bdb
I_-(s,b,\xi)/|1-iU(s,b)|^2} {2\int_0^\infty bdb
I_0(s,b,\xi)/|1-iU(s,b)|^2},\label{xnn}
\end{equation}
where $I_0=1/2(I^\uparrow+I^\downarrow)$ and $I_-=(I^\uparrow-I^\downarrow)$
and $I_0$ obey the sum rule
\[
\int I_0(s,b,\xi) d\xi = \bar n(s,b)Im U(s,b),
\]
here $\bar n(s,b)$ stands for the mean multiplicity in the impact parameter
representation.

With the above relation for the shift in the impact parameter due to the transition flipping quark spin
the following expression for asymmetry $A_N^{\pi^0}$ can be written
\begin{equation} A_N^{\pi^0}(s,\xi)\simeq -x_Fr_{Q}^{flip}\frac{1}{3}\frac{\int_0^\infty bdb
I'_0(s,b,\xi)db/|1-iU(s,b)|^2} {\int_0^\infty bdb
I_0(s,b,\xi)/|1-iU(s,b)|^2},\label{poll}
\end{equation}
where $I'_0(s,b,\xi)={dI_0(s,b,\xi)}/{db}$.  
It is evident that $A_N^{\pi^0}(s,\xi)$
should be positive because $I'_0(s,b,\xi)<0$.

In the model \cite{csn} the function $U(s,b)$ is
chosen  to be  a product of the averaged quark amplitudes
under assumption of the the quasi-independence of valence constituent
quark scattering in the self-consistent mean field. The 
function $U(s,b)$ in a pure imaginary case for the elastic scattering amplitude, which we consider
 here for simplicity has
been written in the following form
\begin{equation} U(s,b) = i\tilde U(s,b)=ig(s)\exp(-Mb/\zeta ),
 \label{x}
\end{equation}
where the factor $g(s)$ increases at large values of $s$ like a power
of energy:
\[
g(s)= \left[1+\alpha\frac{\sqrt{s}}{m_Q}\right]^N,
\]
$M$ is the total mass of $N$ constituent quarks with mass $m_Q$ in
the initial hadrons and parameter $\zeta$  determines  a universal scale for
the quark interaction radius without flipping its spin, i.e. $r_Q=\zeta /m_Q$.

To evaluate asymmetry dependence on $x_F$ and $p_T$
the semiclassical correspondence  between transverse momentum and impact parameter
 has been used.
Integrating by parts, we can rewrite
 the expression for the asymmetry
in the form:
\begin{equation}\label{asym}
A_N^{\pi^0}(s,\xi)\simeq x_Fr_{Q}^{flip}
\frac{M}{3\zeta}\frac{\int_0^\infty bdb
I_0(s,b,\xi)\tilde U(s,b) /[1+\tilde U(s,b)]^3} {\int_0^\infty bdb
I_0(s,b,\xi)/[1+\tilde U(s,b)]^2},
\end{equation}

At small values of $b$  the values of $U$-matrix are large,
and  we can neglect unity in the denominators of the integrands.
Thus the ratio of the two integrals (after integration by parts of nominator in Eq. (\ref{asym}))
 is of order
of unity, i.e.  the energy and $p_T$-independent behavior
of asymmetry $A_N^{\pi^0}$ takes place at the values of transverse momentum $p_T\gg x_F/R(s)$:
\begin{equation} A_N^{\pi^0}(s,\xi)\simeq x_Fr_{Q}^{flip}
\frac{M}{3\zeta}.\label{polllg}
\end{equation}
This  flat transverse momentum dependence of asymmetry results from the similarity of the
rescattering effects for the different spin states, i.e. spin-flip and spin-nonflip
interactions undergo similar absorption at short distances and
the relative magnitude of this absorption does not depend on energy. It can be considered as a one
of the manifestations of the unitarity.
The numerical value of polarization $A_N^{\pi^0}$ can be significant. Indeed, there is
no small factor in (\ref{polllg}).  The function $R(s)$ is the hadron interaction radius ($R(s)\sim \ln s$), the typical numerical
falue of $R$ has been taken to be equal to 1 fm. Thus, the typical value of $x_F/R(s)$ is 0.1 GeV/c and the Eq. (\ref{polllg}) which
is valid at $p_T\gg x_F/R(s)$, should be applicable in the region $p_T>1$ GeV/c.
The value of $r_{Q}^{flip}$ is of order $\sim 10^{-1}$ fm on the basis of the model estimates \cite{ttech,csn,tmf}. 
The radius of quark interaction
$r_{Q}^{flip}$
responsible for the transition $Q_\uparrow\to Q_\downarrow$ changing quark spin.
 The linear increase of asymmetry
with $x_F$ follows from the above considerations which, of course, are approximate and
valid at $x_F$ around unity. Therefore, at smaller values of $x_F$ the linear dependence is distorted.

Thus, Eq. (\ref{polllg} is valid in the region of large $x_F$. 
The flat dependence of asymmetry on
$p_T$ provided by this relation is consistent with the new data from RHIC (we have used data available at the largest values
of $x_F$ due to approximation made in the model and discussed above) \cite{fpio,fpio1} (cf.  Fig. 2,3).
\begin{figure}[htb]
\begin{center}
  \resizebox{10cm}{!}{\includegraphics*{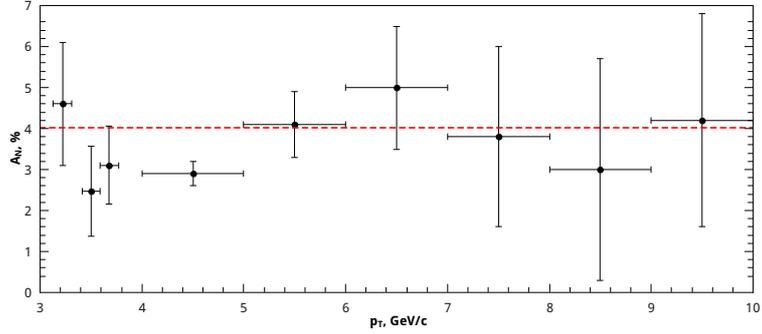}}
\end{center}
\caption{$p_T$ -dependence of the asymmetry $A_N$ in the process $p_\uparrow+p\to\pi^0+X$ at RHIC,
  preliminary data from \cite{fpio,fpio1} correspond to pion isolation of  70 mR, $\sqrt{s}=500$ GeV and $0.32<x_F<0.40$.} \label{tsf}
\end{figure}
\begin{figure}[htb]
\begin{center}
  \resizebox{10cm}{!}{\includegraphics*{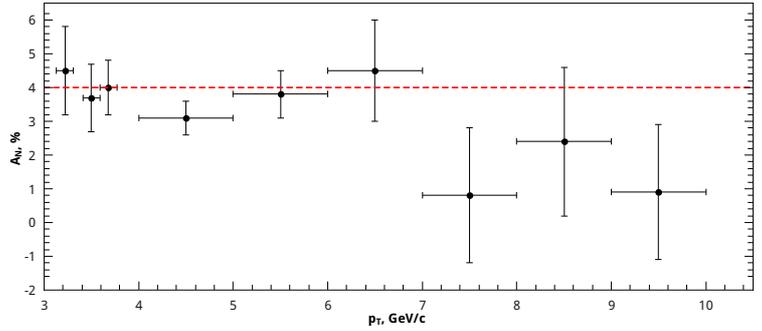}}
\end{center}
\caption{The same plot as in Fig.3, but with
data corresponding to pion isolation of  30 mR.} \label{tsfn}
\end{figure}
Comparison with the data allows one to estimate the value of  $r_{Q}^{flip}$ more precisely , namely
$r_{Q}^{flip}\simeq 0.05$ fm. 
Similar mechanism  generates SSA in the inclusive production of charged pions.
It should be noted that dependencies of SSA on $p_T$ consistent with the flat ones have also been
observed at lower values of $x_F$, namely, in the two regions $0.16<x_F<0.24$ and $0.24<x_F<0.32$ \cite{fpio,fpio1}.
The data demonstrate increase of SSA with $x_F$. Due to limitation of the model for the large $x_F$ region, we have not used those data,
but, in principle, the model is in agreement with them too. To be honest, one should not compare the predictions of the model which are
valid for $x_F$ in the vicinity of unity with the experimental data wich have highest value of $x_F$ around 0.4. However, the data demonstrate
similar flat dependence on transverse momentum in the the rather wide range $x_F$ from 0.16 to 0.4. We therefore  assume that the similar dependence
will be valid at $x_F$ values around unity and perform comparison with the experiment. It should be clarified, that the above agreement with the data has a 
qualitative value only and for the quantitative analysis the data at higher values of $x_F$ are required.
 
The mechanism of chiral quark fluctuation  in the effective field
 with spin flip is
 suppressed compared to the direct elastic
 scattering of quarks  and, therefore, it
should not play a significant role e.g. in the reaction $pp_\uparrow\to pX$ in the fragmentation
 region, but evidently it is not the case for the reaction $pp_\uparrow\to nX$. The above features 
 can be observed experimentally: asymmetry $A_N$ is consistent with zero for proton production
 and significantly deviates from zero for the neutron production in the forward region. 

In ref.  \cite{bsof} an important issue
has been raised, namely
a model when trying to explain spin asymmetries should simultaneously describe
the data for the unpolarized inclusive cross-sections.
In this approach with the effective degrees of freedom --
 constituent quarks and Goldstone bosons -- unpolarized inclusive cross--section at high transverse momenta
 fllows a generic power-like dependencies on $p_T$ .
 At high $p_T$ the power-like dependence $p_T^{-n}$ with $n=6$
takes place. It originates from the singularity at zero impact parameter
 $b=0$.
 The exponent $n$ does not depend
 on $x_F$. This  $p_T^{-6}$--dependence of
the unpolarized inclusive cross--section  is consistent with the respective  data dependence on the transverse momentum (Fig. 5) \cite{ttech}.
\begin{figure}[htb]
\begin{center}
  \resizebox{8cm}{!}{\includegraphics*{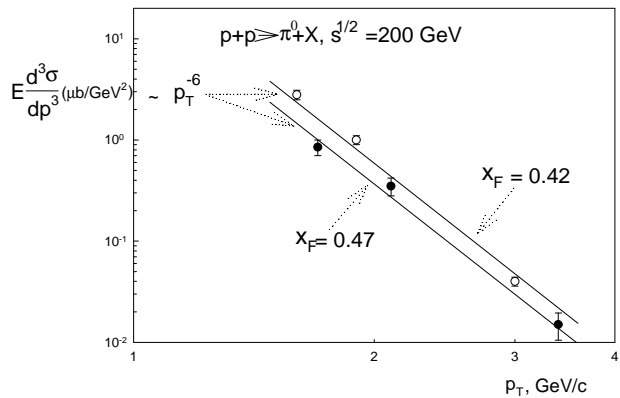}}
\end{center}
\caption{Transverse momentum dependence of unpolarized inclusive cross--section,
experimental data from \cite{star}.} \label{ds}
\end{figure}

\section*{Conclusion}
It was shown that prediction of the spin filtering mechanism \cite{ttech} on the flat $p_T$-dependence of single-spin asymmetries
is consistent with the new experimental data from RHIC. As it often happens, this interpretation 
is not unique. Such flat dependence can  result  from finite size of a constituent
quark and a presence of the orbital angular momentum of the current quarks residing inside the  constituent one \cite{usprd,moch}.
This mechanism is based on the similar ideas as the spin filtering one but is, in principle, different and has no predictive power 
for $x_F$ dependence of SSA.
Thus, we would like to treat the seemingly flat $p_T$-dependence of SSA in favor of the spin filtering mechanism, while the presence of
the internal orbital momentum in the structure of constituent quarks still remains to be an interesting option and cannot
be excluded at the moment (cf. e.g. \cite{arash}).

Finally, we would like to stress again that the experimental data set  \cite{fpio,fpio1} is the preliminary one and we hope that the final  data 
will show smaller  error bars and allow one to provide a quantitative discrimination of the model predictions for SSA.
 
\section*{Acknowledgment}
We are grateful to Yuri Kovchegov for the information on the new experimental data from RHIC and the
interesting discussions. 
\end{document}